\documentclass{ptephy}

\preprintnumber{XXXX-XXXX} 

\usepackage{amsmath} 
\usepackage{graphics} 
\usepackage{bm}

\newcommand{\beq}{\begin{equation}}
\newcommand{\eeq}{\end{equation}}
\newcommand{\bea}{\begin{eqnarray}}
\newcommand{\eea}{\end{eqnarray}}

\begin{document}

\title{Extracting electric dipole breakup cross section of one-neutron
halo nuclei from inclusive breakup observables}

\author{\name{Kazuki Yoshida}{1\ast}, \name{Tokuro Fukui}{1},
\name{Kosho Minomo}{2}\thanks{Present address: Research Center for Nuclear Physics, Osaka University, Ibaraki 567-0047, Japan},
and \name{Kazuyuki Ogata}{1}}

\address{\affil{1}{Research Center for Nuclear Physics, Osaka University, Ibaraki 567-0047, Japan}
\affil{2}{Department of Physics, Kyushu University, Fukuoka 812-8581, Japan}
\email{yoshidak@rcnp.osaka-u.ac.jp}}

\begin{abstract}%
We discuss how to extract an electric dipole (E1) breakup cross section $\sigma ({\rm E1})$ from one-neutron
removal cross sections measured at 250~MeV/nucleon
by using $^{12}$C and $^{208}$Pb targets, $\sigma_{-1n}^{\rm C}$ and
$\sigma_{-1n}^{\rm Pb}$, respectively.
It is shown that within about 5\% error, $\sigma ({\rm E1})$ can be obtained by subtracting
$\Gamma \sigma_{-1n}^{\rm C}$ from $\sigma_{-1n}^{\rm Pb}$, as assumed in preceding studies.
However, for the reaction of weakly-bound projectiles, the scaling factor $\Gamma$ is found to
be about two times as large as that usually adopted.
As a result, we obtain 13--20~\% smaller $\sigma ({\rm E1})$ of $^{31}$Ne at 250~MeV/nucleon
than extracted in a previous analysis of experimental data.
By compiling the values of $\Gamma$ obtained for
several projectiles, $\Gamma=(2.30\pm0.41)\exp(-S_n)+(2.43\pm0.21)$ is obtained, where $S_n$ is
the neutron separation energy.
The target mass number dependence of the
nuclear parts of the one-neutron removal cross section and the elastic breakup cross section is
also investigated.
\end{abstract}

\subjectindex{xxxx, xxx}

\maketitle






\section{Introduction}
\label{s1}

The neutron halo structure~\cite{Tan85,Tan13}, which indicates the breakdown of
the saturation property of the nuclear density,
is one of the novel properties of unstable nuclei.
So far several neutron halo nuclei have been discovered:
$^{11}$Be, $^{15}$C, $^{19}$C, and $^{31}$Ne are well established
one-neutron halo nuclei, and $^6$He, $^{11}$Li, $^{14}$Be, $^{17}$B,
and $^{22}$C are known as two-neutron halo nuclei. Nowadays,
the neutron halo structure is considered to be a rather {\it general}
feature of unstable nuclei far from the stability line.
It is thus important to complete a list
of halo nuclei, which is a hot subject in nuclear physics.

One of the most well known probes for the halo structure is
the interaction cross section $\sigma_{\rm I}$~\cite{Tan85,Tan13,Tan10,Tak12}.
In an experiment, $\sigma_{\rm I}$ are measured for several isotopes with
a target nucleus. A halo nucleus
is identified at a mass number where a large increase in $\sigma_{\rm I}$
is found. Recently, a fully microscopic analysis of $\sigma_{\rm I}$
of Ne isotopes based on the antisymmetrized molecular dynamics (AMD)
wave functions~\cite{Eny12} and the Melbourne nucleon-nucleon
$g$ matrix~\cite{Amo00} was carried out~\cite{Min12}.
It was concluded that $^{31}$Ne is a one-neutron halo nucleus
with a large deformation of the $^{30}$Ne core. A similar analysis
is ongoing for Mg isotopes.

As an alternative probe for the halo structure, it was shown in
Ref.~\cite{Nak09} that the breakup cross section
$\sigma ({\rm E1})$ due to the electric dipole (E1) field
can be utilized; it was shown that for $^{19}$C, a well-known
one-neutron halo nucleus, $\sigma ({\rm E1})$ was indeed large.
This is essentially due to the large cross section
for the soft dipole excitation that is a characteristic of a
halo nucleus.
The authors also obtained a large value of $\sigma ({\rm E1})$ for
$^{31}$Ne, with which $^{31}$Ne was concluded to be a
one-neutron halo nucleus. Since $\sigma ({\rm E1})$ is not an
observable, in the analysis the following
equation was used to obtain it:
\beq
\sigma ({\rm E1}) = \sigma_{-1n}^{\rm Pb} - \Gamma \sigma_{-1n}^{\rm C},
\label{nakamura}
\eeq
where $\sigma_{-1n}^{\rm A}$ is the one-neutron removal cross
section by a target nucleus A and $\Gamma$ is a scaling factor
ranging from 1.7 to 2.6 for the $^{31}$Ne projectile.
However, no quantitative justification
of Eq.~(\ref{nakamura}) for the reaction system was made. Since Eq.~(\ref{nakamura}) is
a key formula in the study of Ref.~\cite{Nak09}, it will be very
important to clarify the validity of the equation.

In this paper, we describe one-neutron removal processes
by means of sophisticated three-body reaction models:
the continuum-discretized coupled-channels method with eikonal
approximation (E-CDCC)~\cite{Oga03,Oga06} for the elastic breakup and the eikonal
reaction theory (ERT)~\cite{Yah11,Has11} for the one-neutron stripping. The purpose
of the present study is
to examine Eq.~(\ref{nakamura}) and find an appropriate value of $\Gamma$.
There exists a number of
works~\cite{ST99,NT99,EB99,Nag01,VE02,EB02,CS02,Mar02,Hus06,Oga09,KM12,KB12}
regarding the assumptions behind Eq.~(\ref{nakamura}),
i.e., interference between nuclear and Coulomb breakup, role of continuum-continuum couplings,
breakup due to the electric quadrupole field, and so forth. Most of them focused on reactions
at relatively lower incident energies, where one may expect the above-mentioned {\lq\lq}higher-order''
effects. In the present study, we consider breakup processes at 250~MeV/nucleon;
at such energies, the mechanism of the breakup reaction is believed to be simple. Nevertheless,
it will be very important to evaluate possible errors of using Eq.~(\ref{nakamura}) quantitatively.
Another important aim of this work is to find a target mass-number ($A$) dependence
of $\sigma_{-1n}$ due to the nuclear interaction, which is essential to determine
the scaling factor $\Gamma$.

The construction of this paper is as follows. In \S\ref{s2}
we briefly recapitulate the formalism of E-CDCC and ERT, and
clarify the condition for Eq.~(\ref{nakamura}) to be satisfied.
In \S\ref{s3} we examine the assumptions behind Eq.~(\ref{nakamura})
one by one.
Then the $A$-dependence of
$\sigma_{-1n}$ due to the nuclear interaction is investigated for
several projectiles and $\Gamma$ is evaluated. We present a functional
form of $\Gamma$ with respect to the neutron separation energy $S_n$.
The $A$-dependence of the nuclear part of the elastic breakup cross section is
also discussed. Finally, a summary is given in \S\ref{s4}.

\section{Formalism}
\label{s2}

\subsection{Three-body system and model space}
\label{s2-1}

\begin{figure}[htbp]
\begin{center}
\includegraphics[width=7cm]{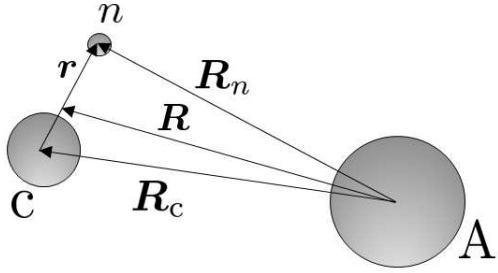}
\caption{Schematic illustration of the ${\rm c}+n+{\rm A}$ three-body system.}
\label{fig1}
\end{center}
\end{figure}

We describe the one-neutron removal process with a ${\rm c}+n+{\rm A}$
three-body system shown in Fig.~\ref{fig1};
c and $n$ are the core nucleus and the valence neutron in
the projectile P, respectively, and A is the target nucleus.
%
The coordinates of P, c, and $n$ relative to A are denoted by ${\bm R}$,
${\bm R}_{\rm c}$, and ${\bm R}_n$, respectively, and ${\bm r}$ represents
the coordinate from c to $n$.
The three-body Schr\"odinger equation is given by
\begin{equation}
\left[  -\frac{\hbar^{2}}{2\mu}{\bm \nabla}_{{\bm R}}^{2}%
+U_{n}(  R_{n})  +U_{\mathrm{c}}(  R_{\mathrm{c}})
+\hat{h}-E\right]  \Psi(  {\bm r},{\bm R})  =0,
\label{sch}%
\end{equation}
where $\mu$ is the reduced mass of the P-A system, $U_{n}$ and $U_{\rm c}$ are,
respectively, the distorting potentials of $n$ and c by A;
 $U_{\rm c}$ consists of the nuclear and Coulomb parts.
$\hat{h}$ is the internal Hamiltonian of P and $E$ is the total energy of
the system.
We solve Eq.~(\ref{sch}) within a model space:
\begin{equation}
\mathfrak{P}\equiv\sum_{i=0}^{i_{\rm max}}
\left\vert i\right\rangle \left\langle i\right\vert
\approx1,
\label{modelspace}%
\end{equation}%
where $\left\vert i\right\rangle$ is the ground state ($i=0$) or
a discretized continuum state ($i>0$) of P.
Equation~(\ref{modelspace}) means that approximately $\mathfrak{P}$
can be regarded as a complete set for describing a reaction process
considered in the present study~\cite{Yah12}.

\subsection{Continuum-discretized coupled-channels method with eikonal
approximation (E-CDCC)}
\label{s2-2}

In E-CDCC~\cite{Oga03,Oga06}, the total wave function $\Psi(  {\bm r},{\bm R})$
is described by
\begin{equation}
\Psi(  {\bm r},{\bm R})  =
\sum_{i}\frac{1}{\sqrt{\hbar v_{i}}}
e^{i(K_{i}z+\eta_i \ln (K_i R-K_i z))}
\psi_{i}(  b,z)
\phi_{i}({\bm r}),
\label{psi-ecdcc}
\end{equation}%
where $\phi_{i}({\bm r})$ is the wave function of P in the
$i$th state satisfying
$\hat{h}\phi_{i}(  {\bm r})  =\varepsilon_{i}\phi_{i}({\bm r})$,
$K_{i}$ ($v_i$) is the relative wave number (velocity) between P and A,
and $\eta_i$ is the Sommerfeld parameter.
$b$ is the impact parameter and $\phi_R$
is the azimuthal angle of ${\bm R}$. The $z$-axis is taken to be the
incident direction.
For simplicity, in Eq.~(\ref{psi-ecdcc}) the $\phi_R$
dependence of the wave function is dropped.
It should be noted that the monopole Coulomb interaction
between P and A is taken into account by using the Coulomb incident wave
function in Eq.~(\ref{psi-ecdcc}).

After solving the E-CDCC equation, Eq.~(4.5) of Ref.~\cite{Yah12},
with the boundary condition
$\lim_{z\rightarrow-\infty}\psi_{i}(  b,z)  =\sqrt{\hbar v_{0}}\delta_{i0}$,
one obtains the eikonal $S$-matrix element
\begin{equation}
S_{i}(  b)  =\frac{1}{\sqrt{\hbar v_{0}}}\lim_{z\rightarrow\infty
}\psi_{i}(  b,z).
\label{s-ecdcc}
\end{equation}%
The elastic breakup cross section $\sigma_{\rm EB}$ is given by
\begin{equation}
\sigma_{\rm EB}=
2\pi\int \sum_{i\neq 0} \left\vert S_{i}(
b)  \right\vert ^{2}bdb.
\label{sigeb-ecdcc}%
\end{equation}

\subsection{Eikonal reaction theory (ERT)}
\label{s2-3}

ERT~\cite{Yah11,Has11} is an extended version of CDCC that is
applicable to the neutron stripping processes, explicitly taking
account of the Coulomb breakup contribution. ERT describes
the total wave function as
\begin{equation}
\Psi(  {\bm r},{\bm R})  =\frac{1}{\sqrt{\hbar\hat
{v}}}
e^{i(\hat{K}z+\hat{\eta} \ln (\hat{K} R-\hat{K} z))}
\Phi(  {\bm r},{\bm R}),
\label{psi-ert}%
\end{equation}%
where the P-A relative wave number is represented by the operator:
\begin{equation}
\hat{K}=\frac{1}{\hbar}\sqrt{2\mu(  E-\hat{h})  }.
\label{opk}%
\end{equation}%
Accordingly,  the P-A Sommerfeld parameter is treated as an operator.
The velocity operator $\hat{v}$ depends on $R$ as
\beq
\hat{v}
=
\sqrt{
\frac{2}{\mu}
\left(
E - \hat{h} - \frac{Z_{\rm P}Z_{\rm A}e^2}{R}
\right)
},
\label{opv}
\eeq
which is the operator form of Eq.~(4.4) of Ref.~\cite{Yah12}
multiplied by $\hbar/\mu$;
$Z_{\rm P}$ $(Z_{\rm A})$ is the atomic number of P (A).
The $S$ matrix operator in ERT is given by
\begin{equation}
\hat{S}\equiv
\exp  \left[ \frac{\mathcal{P}}{i} %
\int_{-\infty}^{\infty}
\!\! \hat{O}^{\dagger}(  z^{\prime})  \left[
U_{n}(  R_{n} ) \! +  U_{\mathrm{c}}(  R_{\mathrm{c}})
\right]  \hat{O}(  z^{\prime})  dz^{\prime}\right],
\label{s-ert}%
\end{equation}%
where
\begin{equation}
\hat{O}(  z)  \equiv\frac{1}{\sqrt{\hbar\hat{v}}}
e^{i(\hat{K}z+\hat{\eta} \ln (\hat{K} R-\hat{K} z))}
\label{opo}%
\end{equation}%
and $\mathcal{P}$ is the path ordering operator with respect to $z$.

Then, the adiabatic approximation $\hat{h} \rightarrow \varepsilon_0$
is made in Eq.~(\ref{s-ert}) to the term related to $U_{n}(  R_{n} )$,
which results in the separation of the $S$-matrix operator:
\begin{equation}
\hat{S}\rightarrow \hat{S}_{n}\hat{S}_{\mathrm{c}}
\label{ssep}
\end{equation}
with
\begin{equation}
\hat{S}_{n}=
\exp\left[  \frac{1}{i\hbar v_{0}}\int_{-\infty}^{\infty
}U_{n}(  R_{n})  dz\right],
\label{sn}
\end{equation}
\begin{equation}
\hat{S}_{\mathrm{c}}=
 \exp\left[\frac{\mathcal{P}}{i} \int_{-\infty
}^{\infty}\hat{O}^{\dagger}(  z)  U_{\mathrm{c}}(
R_{\mathrm{c}})  \hat{O}(  z)  dz\right].
\label{sc}
\end{equation}%
The neutron stripping cross section
$\sigma_{n\mathrm{:STR}}$ is given by
\begin{equation}
\sigma_{n\mathrm{:STR}}=2\pi\int\big\langle0\big\vert\big\vert\hat
{S}_{\mathrm{c}}\big\vert^{2}\big(1-\big\vert\hat{S}_{n}\big\vert^{2}%
\big)\big\vert0\big\rangle bdb.
\label{signstr}
\end{equation}%

\subsection{Assumptions behind the E1 cross section formula}
\label{s2-5}

By definition, $\sigma_{-1n}$ is the sum of $\sigma_{\rm EB}$
and $\sigma_{n\mathrm{:STR}}$:
\begin{equation}
\sigma_{-1n}
=
\sigma_{\rm EB}
+
\sigma_{n\mathrm{:STR}}.
\label{sig1n}
\end{equation}
To extract $\sigma ({\rm E1})$,
first we need the following condition of incoherence between
the nuclear and Coulomb breakup:
\begin{equation}
\sigma_{\rm EB}^{\rm Pb}
\approx
\sigma_{\rm EB(N)}^{\rm Pb}
+
\sigma_{\rm EB(C)}^{\rm Pb},
\label{cond1}
\end{equation}
where we put (N) and (C) to specify the nuclear and Coulomb parts
of $\sigma_{\rm EB}$, respectively.
More explicitly, $\sigma_{\rm EB(N)}$ ($\sigma_{\rm EB(C)}$) is the elastic breakup
cross section evaluated with dropping the off-diagonal coupling potentials due to
the Coulomb (nuclear) interaction in solving Eq.~(\ref{sch}).
The second condition is that the coupled-channel effects caused by the Coulomb
interaction on the neutron stripping is negligibly small:
\begin{equation}
\sigma_{n\mathrm{:STR}}^{\rm Pb}
\approx
\sigma_{n\mathrm{:STR(N)}}^{\rm Pb}.
\label{cond2}
\end{equation}
If Eqs.~(\ref{cond1}) and (\ref{cond2}) are satisfied, we have
\begin{equation}
\sigma_{-1n}^{\rm Pb}
\approx
\sigma_{\rm EB(C)}^{\rm Pb}
+
\sigma_{-1n\mathrm{(N)}}^{\rm Pb},
\label{sig1n2}
\end{equation}
where
\begin{equation}
\sigma_{-1n\mathrm{(N)}}^{\rm Pb}=
\sigma_{\rm EB(N)}^{\rm Pb}+\sigma_{n\mathrm{:STR(N)}}^{\rm Pb}.
\end{equation}
The third condition is given by
\begin{equation}
\sigma_{\rm EB(C)}^{\rm Pb}
\approx
\sigma_{\rm EB(E1)}^{\rm Pb}
\equiv
\sigma ({\rm E1}),
\label{cond3}
\end{equation}
where EB(E1) means the first-order E1 transition cross section.
The fourth and last condition is
\begin{equation}
\sigma_{-1n}^{\rm C}
\approx
\sigma_{-1n\mathrm{(N)}}^{\rm C}.
\label{cond4}
\end{equation}
One may then obtain Eq.~(\ref{nakamura}) with $\Gamma$ defined by
\begin{equation}
\Gamma=
\dfrac{\sigma_{-1n\mathrm{(N)}}^{\rm Pb}}{\sigma_{-1n\mathrm{(N)}}^{\rm C}}.
\label{gamma}
\end{equation}

\section{Results and discussion}
\label{s3}

\subsection{Model setting}
\label{s3-1}

We consider neutron removal processes of projectiles having a
$n$-c structure by $^{12}$C, $^{16}$O, $^{48}$Ca, $^{58}$Ni, $^{90}$Zr,
and $^{208}$Pb at 250~MeV/nucleon.
We take a central Woods-Saxon (WS) potential between the $n$-c pair.
The radius parameter $r_0$ and the
diffuseness parameter $a_0$ together with the $n$-c relative
angular momentum $\ell$ in the ground state, $\ell_0$,
and $S_n$ are shown in Table~\ref{tab1}.
%
\begin{table}[b]
  \caption{Inputs for the $n$-c pair.}
  \begin{center}
  \begin{tabular}{cccccc}
   \hline
    & $r_0$ [fm] & $a_0$ [fm] & \; $\ell_0$ \; & $S_n$ [MeV] & Ref. \\
    \hline
    $^{11}$Be & 1.39 & 0.52 & 0 & 0.503 & \cite{Hus06} \\
    $^{15}$C  & 1.10 & 0.60 & 0 & 1.218 & \cite{Cap12} \\
    $^{19}$C  & 1.25 & 0.70 & 0 & 0.580 & \cite{Kon09} \\
    $^{31}$Ne & 1.25 & 0.75 & 1 & 0.330 & \cite{Hor10} \\
     \hline
    \end{tabular}
    \end{center}
  \label{tab1}
\end{table}
The depth of the WS potential is determined to reproduce $S_n$.
The maximum value $\ell_{\rm max}$ of $\ell$ is set to 3.
For each $\ell$, the continuum state up to $k=0.66$~fm$^{-1}$,
where $k$ is the $n$-c relative wave number,
is discretized by the momentum-bin method with an equal increment
$\Delta k$. We take $\Delta k=0.066$~fm$^{-1}$ for $\ell \neq 0$
and $\Delta k=0.033$~fm$^{-1}$ for $\ell = 0$. The maximum value
of $r$ is set to 200~fm.

The distorting potential $U_{n}$ ($U_{\rm c}$) is evaluated by
a microscopic single (double) folding model;
the Melbourne nucleon-nucleon $g$ matrix~\cite{Amo00} and
the Hartree-Fock (HF)
wave functions of c and A based on the Gogny D1S force~\cite{DG80,Ber91}
are adopted.
This microscopic approach has successfully been applied to
several reaction systems~\cite{Min12,Yah12,Sum12}.
The maximum impact parameter $b_{\rm max}$ is taken to be
50~fm for nuclear breakup processes, whereas we put
$b_{\rm max}=400$ ~fm when Coulomb breakup is included.

\subsection{Examination of the E1 cross section formula}
\label{s3-2}

We show in Table~\ref{tab2}
several cross sections discussed in Sec.~\ref{s2-5}
for the $^{11}$Be, $^{15}$C, $^{19}$C, and $^{31}$Ne projectiles
and the $^{12}$C and $^{208}$Pb targets evaluated by
E-CDCC and ERT.
In Table~\ref{tab2-2} $f_1$, $f_2$, $f_3$, and $f_4$, the errors of the
conditions of Eqs.~(\ref{cond1}), (\ref{cond2}), (\ref{cond3}), and (\ref{cond4}),
respectively, are shown;
$f_i$ ($i=1$--4) is the relative difference between the results on the
left-hand-side and the right-hand-side on each equation.
The value of $\sigma ({\rm E1})$ corresponding to Eqs.~(\ref{nakamura}) and (\ref{gamma})
is evaluated by subtracting $\sigma_{-1n\mathrm{(N)}}^{\rm Pb}$ from $\sigma_{-1n}^{\rm Pb}$.
By taking the relative difference between $\sigma ({\rm E1})$ thus obtained and
$\sigma_{\rm EB(E1)}^{\rm Pb}$ calculated by E-CDCC, we get the total error $f_{\rm tot}$ of
Eq.~(\ref{nakamura}).
%
\begin{table}[t]
    \caption{Cross sections for each projectile (in the unit of mb).}
    \begin{center}
      \begin{tabular}{ccccccccc}
    \hline
    P & $\sigma^{\rm{Pb}}_{\rm{EB}}$ & $\sigma^{\rm{Pb}}_{\rm EB(N)}$ & $\sigma_{\rm EB(C)}^{\rm Pb}$ & $\sigma^{\rm{Pb}}_{\rm EB(E1)}$
      & $\sigma^{\rm{Pb}}_{n\mathrm{:STR}}$ & $\sigma^{\rm{Pb}}_{n\mathrm{:STR(N)}}$
      & $\sigma^{\rm{C}}_{-1n}$ & $\sigma^{\rm{C}}_{-1n\mathrm{(N)}}$
      \\ \hline
    $^{11}$Be & 754 & 108 & 670 & 680 & 364 & 347 & 113 & 110 \\
    $^{15}$C  & 445 & 39  & 418 & 423 & 213 & 196 & 77  &  72 \\
    $^{19}$C  & 769 & 81  & 694 & 701 & 308 & 321 & 99  &  94 \\
    $^{31}$Ne & 812 & 61  & 752 & 758 & 268 & 256 & 87  &  82 \\
    \hline
      \end{tabular}
  \label{tab2}
  \end{center}
\end{table}
%
%
\begin{table}[t]
    \caption{
    Errors of the conditions of Eqs.~(\ref{cond1}), (\ref{cond2}), (\ref{cond3}),
    and (\ref{cond4}) are shown as $f_1$, $f_2$, $f_3$, and $f_4$, respectively.
    $\sigma ({\rm E1})$ evaluated by Eqs.~(\ref{gamma}) and (\ref{nakamura}) is also shown
    in the unit of mb.
    $f_{\rm tot}$ is the total error of Eq.~(\ref{nakamura}). See the text for detail.
    }
    \begin{center}
      \begin{tabular}{ccccccc}
    \hline
    P         & $f_1$ & $f_2$ & $f_3$ & $f_4$ & $\sigma ({\rm E1})$ & $f_{\rm tot}$ \\ \hline
    $^{11}$Be & 3.2\% & 4.6\% & 1.5\% & 2.4\% & 663 & 4.1\% \\
    $^{15}$C  & 2.6\% & 7.9\% & 1.2\% & 6.5\% & 423 & 3.8\% \\
    $^{19}$C  & 0.8\% & 4.2\% & 1.0\% & 5.1\% & 676 & 6.7\% \\
    $^{31}$Ne & 0.1\% & 4.5\% & 0.8\% & 5.9\% & 763 & 1.9\% \\
    \hline
      \end{tabular}
  \label{tab2-2}
  \end{center}
\end{table}
One sees all of the errors are below 8\% for the breakup of these one-neutron halo nuclei
at 250~MeV/nucleon, which validates the use of Eq.~(\ref{nakamura}) for the systems.
On average, $f_1$ and $f_3$ are less than
a few percent, whereas $f_2$, $f_4$, and $f_{\rm tot}$ are about 5\%.

The small $f_1$, i.e., small nuclear and Coulomb interference,
can be understood as follows. First, because of the dominance
of the E1 coupling, the $\ell$ value after the Coulomb breakup
is concentrated to $|\ell_0 \pm 1|$. In fact,
more than 97\% of $\sigma_{\rm EB(C)}$ for $^{11}$Be ($\ell_0=0$)
by $^{208}$Pb comes from the $p$-wave breakup cross section.
On the other hand, there is no such selection for the nuclear breakup;
about 3/4 of $\sigma_{\rm EB(N)}$ goes to $\ell \neq 1$.
Second, as an important aspect of the present study, we discuss the cross section
integrated over the scattering angle. Then there is no interference between different values
of $b$. It is well known that nuclear breakup amplitude is
concentrated at the nuclear surface, whereas the E1 amplitude has
a very long tail with respect to $b$. Therefore, the nuclear and
Coulomb breakup processes occur at different $b$ and populate
different $\ell$ of the breakup state, which results in small
nuclear-Coulomb interference.
It should be noted that if an angular distribution of the breakup cross section
is discussed, because of the coherence of the breakup amplitudes at different $b$,
one can expect non-negligible nuclear-Coulomb interference even at 250~MeV/nucleon.

Equation~(\ref{cond2}) is expected to hold because, as mentioned above,
$\sigma_{n\mathrm{:STR}}$ is due to $U_n(  R_n)$ unless strong coupled-channel
effects caused by the Coulomb interaction exist.
The small $f_2$ obtained will support this picture.
For $f_3$, an important point of the present analysis is that we consider one-neutron
halo nuclei, for which the E2 effective charge $e_{\rm E2}$
is much smaller than the E1 effective charge $e_{\rm E1}$.
Because the E1 coupling strength is small and the scattering
energy is relatively high, one may expect that Eq.~(\ref{cond3})
holds well, which is indeed the case as shown in Table~\ref{tab2-2}.
It should be noted that the very large E1 breakup cross section
by $^{208}$Pb is due to the long-range nature of its amplitude, not to its
strength.
Note also that in the present study deformation effects of c
are neglected; the coupling between the $0^+$ and $2^+$ states of c
can change the E2 transition amplitude.
Inclusion of the core deformation in E-CDCC and
ERT following the recent works~\cite{ML12,Lay14} will be interesting and important
future work.

The conclusions summarized in Table~\ref{tab2-2}
change if we  consider a proton {\lq\lq}halo''
nucleus, e.g., $^8$B. In Table~\ref{tab3} several cross sections
for the $^8$B breakup by $^{208}$Pb at 250~MeV/nucleon
are shown; $r_0=1.25$~fm, $a_0=0.52$~fm,
$\ell_0=1$, and the proton separation energy of 0.137~MeV
are used~\cite{Hus06}.
%
\begin{table}[t]
    \caption{Cross sections for the $^8$B breakup by $^{208}$Pb (in the unit of mb).}
    \begin{center}
  \begin{tabular}{ccccccc}
     \hline
    $\sigma^{\rm{Pb}}_{\rm{EB}}$ & $\sigma^{\rm{Pb}}_{\rm EB(N)}$ & $\sigma_{\rm EB(C)}^{\rm Pb}$&$\sigma^{\rm{Pb}}_{\rm EB(E1)}$ & $\sigma^{\rm{Pb}}_{\rm EB(E2)}$  \\ \hline
    254 & 27 & 258 & 228 & 46
    \\ \hline
  \end{tabular}
      \end{center}
  \label{tab3}
\end{table}
One sees the E2 contribution $\sigma ({\rm E2})$
is about 20\% of $\sigma ({\rm E1})$. This is essentially because $e_{\rm E2}$
of $^8$B is about 2.7 times as large as its $e_{\rm E1}$.
Then, the higher-order effect reduces the sum of $\sigma ({\rm E1})$
and $\sigma ({\rm E2})$ by about 5\%. This somewhat large higher-order
effect is due to the large E2 coupling strength compared with the E1 strength.
In fact, if we perform an all-order calculation including just the
E1 coupling, we obtain 220~mb that is smaller than $\sigma ({\rm E1})$
by only 3\%.
We have thus the addition of $\sigma ({\rm E2})$ to $\sigma ({\rm E1})$
and the decrease in the first-order Coulomb breakup cross section,
$\sigma ({\rm E1})+\sigma ({\rm E2})$,
due to higher-order processes.
In the end,
$\sigma ({\rm E1})$ is smaller than $\sigma_{\rm EB(C)}^{\rm Pb}$ by
about 13\% even at 250~MeV/nucleon.
The nuclear-Coulomb interference of about 12\% also appears for $^8$B
because of the less selectivity of $\ell$.
Therefore, we conclude that we have less validity of Eq.~(\ref{nakamura})
for ${\rm c}+p$ nuclei, even if we consider inclusive breakup observables measured
at 250~MeV/nucleon.
%
%

\subsection{Target mass number dependence of one-neutron removal cross section
due to nuclear interaction}
\label{s3-3}

%
\begin{table}[b]
  \caption{Parameters for the $n$-c pairs.}
  \begin{center}
   \begin{tabular}{cccccccc}
     \hline
 &$^{29}$Ne&  $^{33}$Mg&  $^{35}$Mg&  $^{37}$Mg&  $^{39}$Si&  $^{41}$Si \\ \hline
$\ell_0$& 0&  1&  1&  0 or 1& 1&  1\\
$S_n$ [MeV] &  1.260& 2.640& 1.011&   0.489&  2.080& 0.300
    \\ \hline
    \end{tabular}
    \end{center}
  \label{tab4}
\end{table}
It is shown in Sec.~\ref{s3-2} that at 250~MeV/nucleon Eq.~(\ref{nakamura}) holds well
for one-neutron halo nuclei with $\Gamma$ given by Eq.~(\ref{gamma}).
Before evaluating $\Gamma$, we see the $A$-dependence of
$\sigma_{-1n\mathrm{(N)}}^{\rm A}$. We here consider not only
the four well-established one-neutron halo nuclei listed in Table~\ref{tab2}
but also $^{29}$Ne, $^{33}$Mg, $^{35}$Mg, $^{37}$Mg, $^{39}$Si, and $^{41}$Si.
The newly added six projectiles are expected to have a ${\rm c}+n$ structure
with a (moderately) small value of $S_n$;
the input parameters for them are shown in Table~\ref{tab4}.
We take $r_0=1.20$~fm and $a_0=0.70$~fm for all the systems.
For $^{37}$Mg we assume two possibilities of $\ell_0$, i.e.,
$^{37}$Mg($s$) ($s$-wave) and $^{37}$Mg($p$) ($p$-wave).
The values of $S_n$ for the Mg isotopes are taken from Ref.~\cite{Wat14}
and those for the other nuclei are from Ref.~\cite{Aud03}; for $^{41}$Si
that has a negative mean value of $S_n$~\cite{Aud03} we use $S_n=0.3$~MeV.

\begin{figure}[t]
\begin{center}
\includegraphics[width=\textwidth]{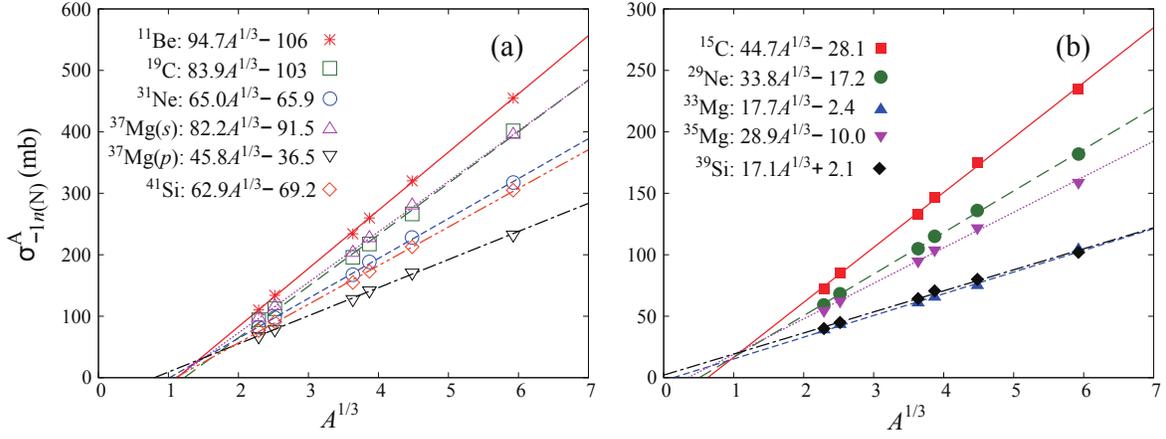}
\caption{
$\sigma_{-1n\mathrm{(N)}}^{\rm A}$
as a function of $A^{1/3}$. Panels (a) and (b) correspond to the
projectiles having $S_n < 1.0$~MeV and $S_n > 1.0$~MeV, respectively.}
\label{fig2}
\end{center}
\end{figure}

We show in Fig.~\ref{fig2} $\sigma_{-1n\mathrm{(N)}}^{\rm A}$
as a function of $A^{1/3}$.
%
Panels (a) and (b) correspond to the
projectiles having $S_n$ smaller and larger than 1~MeV, respectively.
Clearly $\sigma_{-1n\mathrm{(N)}}^{\rm A}$ follows the scaling law
of $A^{1/3}$ for all the projectiles. In each panel the result
of a fitting by $a_{\rm P} A^{1/3} + b_{\rm P}$ is given.
It should be noted that $a_{\rm P}$ and $b_{\rm P}$
have a rather strong dependence on P.

The success of the $a_{\rm P} A^{1/3} + b_{\rm P}$ scaling
of $\sigma_{-1n\mathrm{(N)}}^{\rm A}$ suggests
\begin{equation}
\Gamma=
\dfrac{\bar{a}_{\rm P} R_{\rm Pb} + b_{\rm P}}
{\bar{a}_{\rm P} R_{\rm C} + b_{\rm P}}
=
\dfrac{R_{\rm Pb} + \bar{b}_{\rm P}}
{R_{\rm C} + \bar{b}_{\rm P}},
\end{equation}
where $R_{\rm A}$ is the radius of the nucleus A,
$\bar{a}_{\rm P} \sim 1.2 a_{\rm P}$, and
$\bar{b}_{\rm P}\equiv b_{\rm P}/\bar{a}_{\rm P}$.
Apparently $\bar{b}_{\rm P}$ is related to an effective radius of P,
which naively suggests $0 \le \bar{b}_{\rm P} \le R_{\rm P}$.
In fact, $\Gamma$ is considered in Ref.~\cite{Nak09} to be in the range of
\begin{equation}
\dfrac{R_{\rm Pb} + R_{\rm P}}
{R_{\rm C} + R_{\rm P}}
\le
\Gamma
\le
\dfrac{R_{\rm Pb}}
{R_{\rm C}}.
\label{gammank}
\end{equation}
The lower limit corresponds to the strong absorption limit and
the upper limit to the picture of the Serber model~\cite{Ser47}.
Our present calculation suggests, however, that $\bar{b}_{\rm P}$ can be
negative, which results in $\Gamma$ larger than the upper limit
of Eq.~(\ref{gammank}).

%
\begin{table}[b]
  \caption{The scaling factor $\Gamma$ for the projectiles.}
 \begin{center}
  \begin{tabular}{cccccccccccc}
     \hline
  &$^{11}$Be& $^{15}$C& $^{19}$C& $^{29}$Ne&  $^{31}$Ne&  $^{33}$Mg&  $^{35}$Mg&  $^{37}$Mg($s$)&  $^{37}$Mg($p$)&  $^{39}$Si&  $^{41}$Si \\
\hline
$\Gamma$ & 4.12 & 3.26 & 4.26 & 3.07 & 3.86 & 2.68 & 2.91 & 4.05 & 3.45 & 2.56 & 3.98 \\ \hline
  \end{tabular}
\end{center}
\label{tab4-2}
\end{table}

\begin{figure}[t]
\begin{center}
\includegraphics[width=0.6\textwidth]{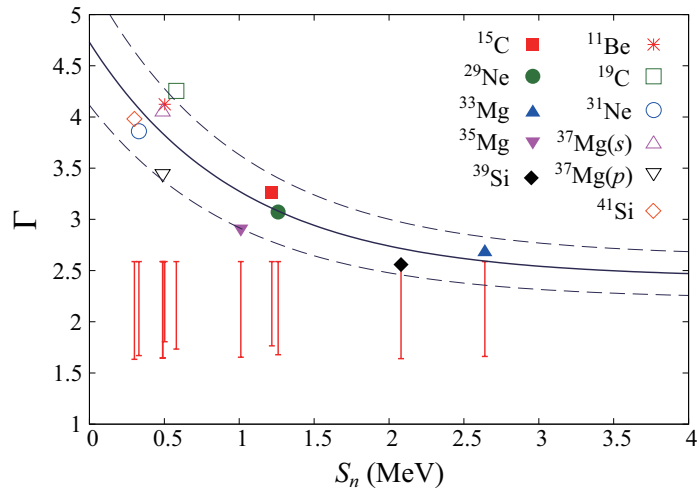}
\caption{
Plot of $\Gamma$ as a function of $S_n$.
The vertical bars show the range of $\Gamma$ given by Eq.~(\ref{gammank}).
The result of a functional fit of Eq.~(\ref{gfit}) is also shown.
}
\label{fig3}
\end{center}
\end{figure}

The results of $\Gamma$ are shown in Table~\ref{tab4-2} and
plotted in Fig.~\ref{fig3} as a function of $S_n$;
in the figure the range of $\Gamma$ assumed in Ref.~\cite{Nak09}, Eq.~(\ref{gammank}),
is shown by a bar for each P.
One sees that $\Gamma$ is located around 4 when $S_n < 1$~MeV and
decreases as $S_n$ increases, toward the upper limit of Eq.~(\ref{gammank}).
%
It is found that $\Gamma$ is well fitted by
\beq
\Gamma=
(2.30\pm0.41)e^{-S_n} + (2.43\pm0.21),
\label{gfit}
\eeq
which is shown by the solid and dashed curves in Fig.~\ref{fig3}. This
simple functional form can be helpful for practical use.

As shown in Fig.~\ref{fig3}, for $^{31}$Ne (open circle) the result of $\Gamma$ of the present study
is almost two times as large as the mean value adopted in the previous study~\cite{Nak09}.
Consequently, we obtain about
13\% (20\%) smaller $\sigma ({\rm E1})$ than that evaluated with the
maximum (minimum) value of $\Gamma$, 2.6 (1.7), of Eq.~(\ref{gammank}).
It should be noted that in the present study we do not consider a spectroscopic
factor ${\cal S}$; we assume ${\cal S}=1$ for all the projectiles.
One may obtain ${\cal S}$ by comparing the theoretical cross section
with experimental data, as in Refs.~\cite{Yah11,Hor10}. In this study,
however, we focus on the $A$-dependence of the cross sections and
the values of $\Gamma$.
It is rather obvious that ${\cal S}$ has very small effect on them.
To be accurate, the one-neutron removal process for
a projectile having a loosely-bound neutron is peripheral.
Thus, what is to be determined through the reaction analysis is not ${\cal S}$
but the asymptotic normalization coefficient (ANC). As shown in Fig.~1
of Ref.~\cite{Yah11}, the ANC evaluated by the experimental data~\cite{Nak09} has
very weak $A$-dependence. Therefore, even though a rather naive structural model
of the projectiles is adopted in this study, the conclusions drawn
above are expected to be quite robust.

\subsection{Scaling of nuclear elastic breakup cross section}
\label{s3-4}

%
\begin{table}[b]
  \caption{Fitting parameters of the effective radius and width of $\sigma_{\rm EB(N)}$ given in the
  unit of fm. $c_2 / c_1$ shows the importance of the $A^{2/3}$ dependence compared to the $A^{1/3}$
  dependence.
  $\sigma_{\rm EB}^{\rm C}$ and $\sigma_{\rm EB(N)}^{\rm C}$ (in the unit of mb)
    and its ratio $\sigma_{\rm EB(N)}^{\rm C}/\sigma_{\rm EB}^{\rm C}$ are also shown.
    See the text for details.
    }
 \begin{center}
 \begin{tabular}{cccccccccc}
 \hline
 & $\alpha _{R}$ &$\beta _{R}$ & $\alpha_{D}\times100$ &$\beta_{D}\times100$ & $c_{2}/c_{1}$ & \; &$\sigma^{\rm{C}}_{\rm{EB}}$ & $\sigma^{\rm{C}}_{\rm EB(N)}$ & $\sigma_{\rm EB(N)}^{\rm C}/\sigma_{\rm EB}^{\rm C}$ \\ \hline
$^{11}$Be & 1.79 & 1.13 & 3.03 & $-2.76$ & $-3.58$ & & 15.13 & 12.67 & 0.837 \\
$^{15}$C & 1.41 & 3.33 & 1.16 & $-1.43$ & 0.886 & & 7.41 & 4.59 & 0.620  \\
$^{19}$C & 1.70 & 2.54 & 2.26 & $-2.81$ & 3.99 & & 12.01 & 8.77 & 0.731  \\
$^{29}$Ne & 1.29 & 4.50 & 0.980 & $-1.39$ & 0.483 & & 6.27 & 3.78 & 0.599  \\
$^{31}$Ne & 1.61 & 3.91 & 1.67& $-2.65$ & 1.19 & & 9.32 & 5.58 & 0.603  \\
$^{33}$Mg & 1.39 & 4.20 & 0.270 & $-0.240$ & 0.469 & & 2.75 & 1.44 & 0.525  \\
$^{35}$Mg & 1.37 & 4.75 & 0.700 & $-0.990$ & 0.487 & & 5.11 & 2.77 & 0.542  \\
$^{37}$Mg($s$) & 1.37 & 4.73 & 2.70 & $-4.59$ & 0.571 & & 13.15 & 8.21 & 0.624  \\
$^{37}$Mg($p$) & 1.28 & 5.21 & 1.30 & $-2.20$ & 0.421 & & 7.16 & 4.03 & 0.563  \\
$^{39}$Si & 1.23 & 5.00 & 0.360 & $-0.430$ & 0.348 & & 3.32 & 1.71 & 0.514  \\
$^{41}$Si & 1.31& 5.20 & 1.83 & $-3.34$ & 0.466 & & 9.48 & 5.03 & 0.530  \\
\hline
\end{tabular}
\end{center}
\label{tab56}
\end{table}
In this subsection we discuss the $A$-dependence of
$\sigma_{\rm EB(N)}$, considering a possibility of extracting $\sigma ({\rm E1})$
from {\it exclusive} breakup observables~\cite{Fuk04,Aum99,Pal03}.
Though it is widely believed that $\sigma_{\rm EB(N)}$ follows
the $A^{1/3}$ scaling~\cite{Hus06}, there exist several works
that report different results~\cite{Nag01,Oga09,Has11}.
In this paper we follow the prescription of Ref.~\cite{Has11}. First, we determine
the effective radius $R_{\rm EB}$ from the peak of the integrand of Eq.~(\ref{sigeb-ecdcc})
divided by $b$.
Then the effective width $D_{\rm EB}$ is evaluated by
$D_{\rm EB}=\sigma_{\rm EB(N)}/(2\pi R_{\rm EB})$.
By looking at the $A$-dependence of $R_{\rm EB}$ and $D_{\rm EB}$,
one can see the $A$-dependence of $\sigma_{\rm EB(N)}$.

\begin{figure}[t]
\begin{center}
\includegraphics[width=\textwidth]{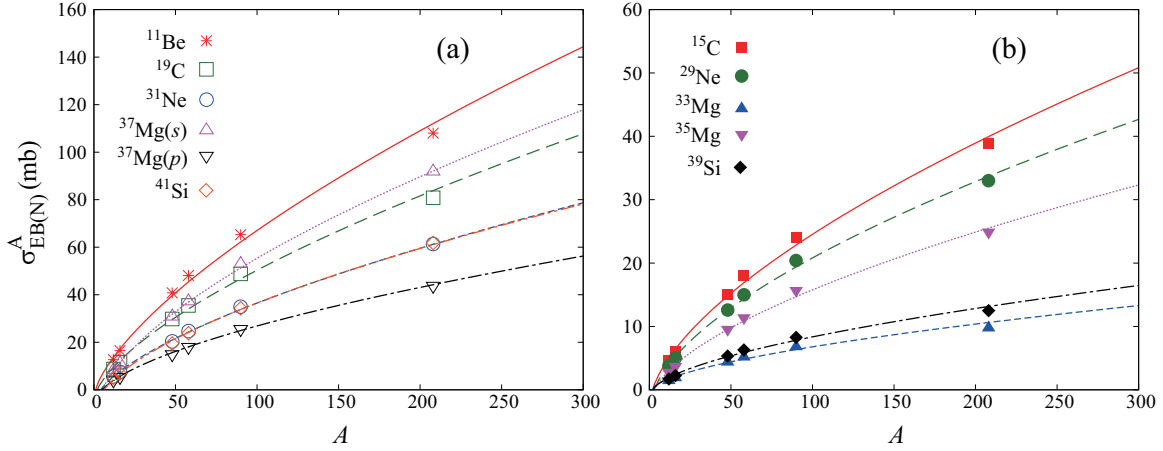}
\caption{Same as in Fig.~\ref{fig2} but for $\sigma_{\rm EB(N)}$ and plotted as a function of $A$.}
\label{fig4}
\end{center}
\end{figure}

It is found, as in Ref.~\cite{Has11}, that both $R_{\rm EB}$ and
$D_{\rm EB}$ are fitted well by the $\alpha_{X}A^{1/3}+\beta_{X}$
form ($X$ is $R$ or $D$). The values of $\alpha_{R}$, $\beta_{R}$,
$\alpha_{D}$, $\beta_{D}$
are given in Table~\ref{tab56}. The resulting functional form of
$\sigma_{\rm EB(N)}$ is
\beq
\sigma_{\rm EB(N)}
=
2\pi (\alpha_{R}A^{1/3}+\beta_{R})
(\alpha_{D}A^{1/3}+\beta_{D})
\equiv
c_{2} A^{2/3} + c_{1} A^{1/3} + c_{0}.
\label{ebnfit}
\eeq
In Fig.~\ref{fig4} we compare Eq.~(\ref{ebnfit}) with
$\sigma_{\rm EB(N)}$ obtained by E-CDCC for the six target nuclei,
as in Fig.~\ref{fig2}. One sees clearly that Eq.~(\ref{ebnfit})
works well for all the projectiles. In the sixth column of
Table~\ref{tab56}, we show $c_{2}/c_{1}$ that gives a deviation from
the $A^{1/3}$ scaling formula. In all the cases $c_{2}/c_{1}$ is
larger than several tens of percent. For $^{11}$Be and $^{19}$C
we have $|c_{2}/c_{1}|\sim4$, which indicates the dominance of the
$A^{2/3}$ dependence.
%

One of the key ingredients for the $A$-dependence of $\sigma_{\rm EB(N)}$
is the c-A distorting potential, which is microscopically
calculated in the present study. Note that the microscopic double folding
model used has successfully been applied to several reaction processes at
around 250~MeV/nucleon~\cite{Yah12,Sum12}.
If we adopt the parameters for the $^{10}$Be-A system given in Table~I
of Ref.~\cite{Hus06}, we obtain a similar result to that of Ref.~\cite{Hus06},
i.e., the $A^{1/3}$ scaling. Another important aspect of
the present study is the incident energy, i.e., 250~MeV/nucleon.
If we evaluate $\sigma_{\rm EB(N)}$ of $^{11}$Be at 70~MeV/nucleon
by E-CDCC with microscopic distorting potentials calculated at the energy,
we have an $A$-dependence slightly weaker than $A^{1/3}$; this is
consistent with the result of Ref.~\cite{Oga09}.
Further investigation will be necessary to draw a definite conclusion
on the $A$-dependence
of $\sigma_{\rm EB(N)}$. At this stage, it is difficult to
find a simple formula to extract $\sigma ({\rm E1})$ from
exclusive breakup observables, i.e.,
$\sigma_{\rm EB}^{\rm Pb}$ and $\sigma_{\rm EB}^{\rm C}$.

Additionally, we remark the importance of Coulomb breakup
by a $^{12}$C target.
We show $\sigma_{\rm EB}^{\rm C}$, $\sigma_{\rm EB(N)}^{\rm C}$, and
the ratio $\sigma_{\rm EB(N)}^{\rm C}/\sigma_{\rm EB}^{\rm C}$
in Table~\ref{tab56}.
One sees that $\sigma_{\rm EB(N)}^{\rm C}/\sigma_{\rm EB}^{\rm C}$ is
considerably smaller than unity. Thus, we need to consider
the contribution from Coulomb breakup of several tens of percent,
when we consider elastic breakup processes
of a projectile consisting of a core nucleus and a loosely-bound
neutron by $^{12}$C at 250~MeV/nucleon.

\section{Summary}
\label{s4}
We have examined the E1 cross section formula, Eq.~(\ref{nakamura}),
by describing the one-neutron removal process at 250~MeV/nucleon
with three-body reaction models. The elastic breakup and the one-neutron
stripping are described by
the continuum-discretized coupled-channels method with the eikonal
approximation (E-CDCC) and the
eikonal reaction theory (ERT), respectively.
We took $^{11}$Be, $^{15}$C, $^{19}$C, $^{29}$Ne, $^{31}$Ne, $^{33}$Mg,
$^{35}$Mg, $^{37}$Mg, $^{39}$Si, and $^{41}$Si
for the projectile, and $^{12}$C, $^{16}$O, $^{48}$Ca, $^{58}$Ni, $^{90}$Zr,
and $^{208}$Pb for the target nucleus.

Four conditions behind Eq.~(\ref{nakamura}) are clarified
and validated one by one within about 5\% error for the breakup of one-neutron halo projectiles
at 250~MeV/nucleon.
The scaling factor $\Gamma$ is defined by the ratio of the
one-neutron removal cross section for the $^{208}$Pb target due to
nuclear interactions, $\sigma_{-1n\mathrm{(N)}}^{\rm Pb}$, to
that for the $^{12}$C target, $\sigma_{-1n\mathrm{(N)}}^{\rm C}$.
It is found that $\sigma_{-1n\mathrm{(N)}}$ follows the $aA^{1/3}+b$ form, where
$A$ is the target mass number, as assumed in preceding studies.
The constant $b$ of the formula, however,
is shown to be negative for almost all the projectiles considered.
This gives somewhat large enhancement of $\Gamma$. We obtained $\Gamma$ for $^{31}$Ne
that is about two times as large as the mean value used in the previous analysis.
Consequently, the E1 cross section of $^{31}$Ne is reduced by 13--20~\%.
We have found the following functional form of $\Gamma$:
$\Gamma=(2.30\pm0.41)\exp(-S_n)+(2.43\pm0.21)$ with $S_n$
the neutron separation energy.

The $A$-dependence of the nuclear elastic-breakup cross section $\sigma_{\mathrm{EB(N)}}$
is also investigated. It is found that $\sigma_{\mathrm{EB(N)}}$
follows $c_2 A^{2/3} + c_1 A^{1/3} +c_0$, i.e., mixture of the $A^{2/3}$ and $A^{1/3}$
scaling. Furthermore, contribution of the Coulomb breakup of several tens of percent,
which is often neglected, is clarified in the breakup
of projectiles having a loosely-bound neutron by the $^{12}$C target.
At this stage, it is quite difficult to find a simple formula to
extract the E1 cross section from exclusive elastic breakup observables.

\section*{Acknowledgment}

The authors thank M.~Yahiro for valuable comments on this study,
and M.~Kimura and S.~Watanabe for providing information on Mg isotopes.
This research was supported in part by Grant-in-Aid of the Japan
Society for the Promotion of Science (JSPS).

\end{document}